\documentclass[11pt]{article} 

\usepackage[utf8]{inputenc} 

\usepackage{geometry} 
\usepackage{graphicx,psfrag} 
\usepackage{color}

\definecolor{Red}{rgb}{1,0,0}
\definecolor{Blue}{rgb}{0,0,1}
\definecolor{Olive}{rgb}{0.41,0.55,0.13}
\definecolor{Green}{rgb}{0,1,0}
\definecolor{MGreen}{rgb}{0,0.8,0}
\definecolor{DGreen}{rgb}{0,0.55,0}
\definecolor{Yellow}{rgb}{1,1,0}
\definecolor{Cyan}{rgb}{0,1,1}
\definecolor{Magenta}{rgb}{1,0,1}
\definecolor{Orange}{rgb}{1,.5,0}
\definecolor{Violet}{rgb}{.5,0,.5}
\definecolor{Purple}{rgb}{.75,0,.25}
\definecolor{Brown}{rgb}{.75,.5,.25}
\definecolor{Grey}{rgb}{.5,.5,.5}

\usepackage{booktabs} 
\usepackage{array} 
\usepackage{paralist} 
\usepackage{verbatim} 
\usepackage{subfigure} 
\usepackage{amsmath,amssymb,amsthm}

\usepackage{fancyhdr} 
\usepackage{sectsty}
\allsectionsfont{\sffamily\mdseries\upshape} 

\theoremstyle{plain}

\newtheorem{theorem}{Theorem} 
 
\newtheorem{corollary}{Corollary}

\newtheorem{lemma}{Lemma}

\newtheorem{bound}{Bound}

\theoremstyle{remark}

\newtheorem{remark}{Remark}

\theoremstyle{definition}

\newcommand{\p}{{\rm P}}

\def\cS{{\cal S}}

\def\cW{{\cal W}}
\def\cX{{\cal X}}

\title{An information inequality and evaluation of Marton's inner bound for binary input broadcast channels}
\author{Chandra Nair \and Zizhou ``Vincent" Wang \and Yanlin Geng }

\begin{document}
\maketitle

\begin{abstract}
We establish an information inequality that is intimately connected to the evaluation of the 
sum rate given by Marton's inner bound for two receiver broadcast channels with a binary input alphabet. This generalizes a recent result where the inequality was established for a particular channel, the binary skew-symmetric broadcast channel. The inequality implies that randomized time-division strategy indeed achieves the sum rate of Marton's inner bound for all binary input broadcast channels.
\end{abstract}

\section{Introduction}
A two-receiver broadcast channel models the communication scenario where two (independent) messages are to be transmitted from a sender $X$ to two receivers $Y,Z$. Each receiver is interested in decoding his/her message. A transition probability matrix given by $p(y,z|x)$ models the stochastic nature of the errors introduced during the communication. For formal definitions and early results the reader can refer to $\cite{cov72,cov98}$. 

\subsection{Background}
The following region obtained by Marton\cite{mar79} represents the best-known achievable region to-date:
\begin{bound} \cite{mar79}
\label{bd:inner}
The set of rate-pairs $(R_1,R_2)$ satisfying the following constraints:
\begin{align*}
R_1 &\leq I(U,W;Y) \\
R_2 &\leq I(V,W;Z) \\
R_1 + R_2 &\leq \min \{I(W;Y), I(W;Z)\} + I(U;Y|W) + I(V;Z|W) - I(U;V|W)
\end{align*}
for any set of random variables $(U,V,W)$ such that $(U,V,W)\to X \to (Y,Z)$ forms a Markov chain are achievable.
\end{bound}

Recently Gohari and Ananthram\cite{goa09} used a remarkable perturbation-based argument to establish that it suffices to consider $(U,V,W)$ with alphabet sizes bounded by $|U| \leq |X|, |V| \leq |X|, |W| \leq |X|+4$ to compute the extreme points of Bound \ref{bd:inner}. In general the computation of Marton's inner bound is difficult, and prior to \cite{goa09}, this bound was not strictly evaluatable. Even with these bounds on cardinalities, explicit evaluation of the bounds is still a difficult task. 

The following region represents an outer-bound to the capacity region of the broadcast channel.
\begin{bound} \cite{nae07}
\label{bd:outer}
The union of rate-pairs $(R_1,R_2)$ satisfying the following constraints:
\begin{align*}
R_1 &\leq I(U;Y) \\
R_2 &\leq I(V;Z) \\
R_1 + R_2 &\leq  I(U;Y) + I(V;Z|U)  \\
R_1 + R_2 &\leq  I(V;Z) + I(U;Y|V)
\end{align*}
over all pairs of random variables $(U,V)$ such that $(U,V)\to X \to (Y,Z)$ forms a Markov chain forms an outer-bound to the capacity region of the broadcast channel.
\end{bound}

The capacity regions of special classes of broadcast channels have been established and in every case it turns out that Bounds \ref{bd:inner} and \ref{bd:outer} agree. In order to study whether the Bounds \ref{bd:inner} and \ref{bd:outer} are indeed different or whether they are different representations of the same region, the authors \cite{naw08} studied a particular channel called the binary skew-symmetric broadcast channel (BSSC). The authors conjectured that for BSSC the following inequality holds:
\begin{equation}
\label{eq:bssc}
I(U;Y) + I(V;Z) - I(U;V) \leq \max\{I(X;Y), I(X;Z)\}.
\end{equation}

The authors further showed that, assuming \eqref{eq:bssc} holds, the Bounds \ref{bd:inner} and \ref{bd:outer} {\em differed} for BSSC. 

In \cite{goa09}, the authors established that Bounds \ref{bd:inner} and \ref{bd:outer} were indeed different for BSSC without actually establishing that \eqref{eq:bssc} was true. They verified that \eqref{eq:bssc} was indeed plausible by confirming it for  a large number of (randomly-generated) samples from the cardinality constrained space.

In \cite{jon09} the validity of the inequality \eqref{eq:bssc} was established rigorously using a modification of the perturbation-based arguments\cite{goa09}. Further the authors\cite{jon09} also established that in order to compute the maximum sum-rate for Marton's inner bound it suffices to consider $|W| \leq |X|, |U| \leq |X|, |V| \leq |X|$, a mild improvement over the results of \cite{goa09} for the sum-rate computation. Further this result also quantifies the gap between the sum-rate estimates given by the inner and outer bounds for the BSSC.

\subsection{Summary of results}

The main result of the paper is the following:

\begin{theorem}
\label{th:main0}
Consider a five tuple of random variables  $(U,V,X,Y,Z)$ such that $(U,V) \to X \to (Y,Z)$ forms a Markov chain and further let $|X|=2$. Then the following inequality holds:
\begin{equation}I(U;Y) + I(V;Z) - I(U;V) \leq \max \{I(X;Y), I(X;Z)\}. 
\label{eq:main}
\end{equation}
\end{theorem}

This generalizes \eqref{eq:bssc} to be true for {\em every} binary-input broadcast channel.
Combining this result with the cardinality bounds for the sum-rate obtained in \cite{jon09}, we also establish that the maximum sum rate given by Marton's coding strategy indeed matches that given via the randomized time-division strategy\cite{nae07}, a much simpler achievable strategy for any binary input broadcast channel.

\begin{corollary}
The maximum value of the sum-rate for Marton's inner bound for any binary-input broadcast channel is given by
\begin{equation*}
\max_{p(w,x)}
\min \{ I(W;Y), I(W;Z)\} +
P\{W=0\}I(X;Y|W=0)+ P\{W=1\}I(X;Z|W=1)
\end{equation*}
where  $|W|=2$.
\label{co:mib}
\end{corollary}

\subsubsection{Randomized time-division strategy}

 Randomized time-division (R-TD) strategy\cite{nae07} corresponds to an achievable strategy for the following setting of $(U,V,W)$ in Bound \ref{bd:inner}: $W=0$ implies that $U=X, V = \emptyset$; and $W=1$ implies that $V=X, U =\emptyset$ (where $\emptyset$ refers to the trivial random variable). Observe that this corresponds to a time-division strategy except that the slots for which communication occurs to one receiver  is also drawn from a codebook which conveys additional information. 

\subsubsection{Relationship between Theorem \ref{th:main0} and $\Gamma_5^*$}

Recently there has been a lot of interest in information inequalities and the study of the structure of the entropic space $\Gamma_N^*$.   Theorem \ref{th:main0} refers to a subset, $\cS$, of points in $\Gamma_5^*$: those corresponding to a five tuple of random variables  $(U,V,X,Y,Z)$ such that $(U,V) \to X \to (Y,Z)$ forms a Markov chain and with a binary constraint on the cardinality of $X$, i.e. $|X|=2$.
It shows that the points in $\cS$ have to lie in the union of two half-spaces induced by the two hyperplanes:
\begin{align*}
I(U;Y) + I(V;Z) - I(U;V) & \leq I(X;Y)\\
I(U;Y) + I(V;Z) - I(U;V) & \leq I(X;Z).
\end{align*}
Since the inequalities are tight, $\cS$ is not a convex region in general. The non-convexity of the region also gives a heuristic reasoning as to why Shannon-type inequalities may not be sufficient to establish Theorem \ref{th:main0}.

Before we go into the proof, we will show how Corollary \ref{co:mib} follows from Theorem \ref{th:main0}.

\section{Proof of Corollary \ref{co:mib}}

We reproduce the following lemma(Claim 4, section 3.1) from \cite{jon09}.

\begin{lemma}\cite{jon09}
For a discrete memoryless broadcast channel,
to compute the maximum of 
$$ \lambda I(W;Y)+ (1-\lambda)I(W;Z) + I(U;Y|W) +
I(V;Z|W) - I(U;V|W), 0 \leq \lambda \leq 1 $$ over all choices of
$(U,V,W) \to X \to (Y,Z)$
it suffices to restrict to $|\cW| = |\cX|$.
\end{lemma}

Hence it follows that to evaluate the Marton's sum-rate for binary input broadcast channel it suffices to look at $|\cW|=2$.

\medskip

Thus we need to show that the maximum sum-rate $\bar R$ obtained by the randomized time-division strategy indeed matches the maximum sum rate $R$  given by Marton's inner bound.

\begin{proof}
Clearly, we have $R \geq \bar R$ as $\bar R$ is a restriction of the choice of $U,V,W$.

Consider a $U,V,W$ that achieves the maximum sum-rate  $R$. We consider two cases:

{\em Case 1:} 
\begin{align*}& I(X;Y|W=0) \geq I(X;Z|W=0) ~ \mbox{and} ~ I(X;Y|W=1) \geq
I(X;Z|W=1), ~\mbox{or}~ \\
& I(X;Z|W=0) \geq I(X;Y|W=0)~ \mbox{and}~ I(X;Z|W=1)
\geq I(X;Y|W=1).
\end{align*}

W.l.o.g. say the former holds, i.e. 
\begin{equation}I(X;Y|W=0) \geq I(X;Z|W=0) ~ \mbox{and} ~ I(X;Y|W=1) \geq
I(X;Z|W=1).
\label{eq:eqc1}
\end{equation}
Clearly
\begin{align*}
R & = \min \{I(W;Y), I(W;Z)\} + I(U;Y|W) + I(V;Z|W) - I(U;V|W) \\
&= \min \{I(W;Y), I(W;Z)\} + \p(W=0) \big(I(U;Y|W=0) + I(V;Z|W=0) - I(U;V|W=0) \big) \\
& \quad + \p(W=1) \big(I(U;Y|W=1) + I(V;Z|W=1) - I(U;V|W=1) \big) \\
& \stackrel{(a)}{\leq} \min \{I(W;Y), I(W;Z)\} + \p(W=0) I(X;Y|W=0) + \p(W=1) I(X;Y|W=1) \\
& \leq \min \{I(W;Y), I(W;Z)\} + I(X;Y|W) \leq I(X;Y) \leq \bar R,
\end{align*}
where $(a)$ follows from Theorem \ref{th:main0} and \eqref{eq:eqc1}.

\medskip

{\em Case 2:} 
\begin{equation} I(X;Y|W=0) \geq I(X;Z|W=0) ~ \mbox{and} ~ I(X;Z|W=1) \geq
I(X;Y|W=1).
\label{eq:eqc2}
\end{equation}

Observe that
\begin{align*}
R & = \min \{I(W;Y), I(W;Z)\} + I(U;Y|W) + I(V;Z|W) - I(U;V|W) \\
&= \min \{I(W;Y), I(W;Z)\} + \p(W=0) \big(I(U;Y|W=0) + I(V;Z|W=0) - I(U;V|W=0) \big) \\
& \quad + \p(W=1) \big(I(U;Y|W=1) + I(V;Z|W=1) - I(U;V|W=1) \big) \\
& \stackrel{(a)}{\leq} \min \{I(W;Y), I(W;Z)\} + \p(W=0) I(X;Y|W=0) + \p(W=1) I(X;Z|W=1)  \leq \bar R,\end{align*}
where $(a)$ follows from Theorem \ref{th:main0} and \eqref{eq:eqc2}.

This implies $R \leq \bar R$ and thus we complete the proof of Corollary \ref{co:mib}.\end{proof}

\section{Proof of Theorem \ref{th:main0}}

The idea of the proof is to fix a $p(y,z|x)$ (i.e. a particular broadcast channel) and show that for all $p_o(x)$ we have that
$$ \max_{p(u,v,x): p(x)=p_o(x)} I(U;Y) + I(V;Z) - I(U;V) \leq \max \{ I(X;Y), I(X;Z) \}. $$

Denote $LHS$ and $RHS$ as the left-hand side and right-hand side of the
inequality~(\ref{eq:main}), respectively. Let $p_{uv}=\p (U=u,V=v)$. Also
we use the following notation: $U\land V$ (and), $U\lor V$ (or), $U\oplus V$
(xor), $\bar{U}$ (not). 

\begin{remark}
\label{re:suff}
From \cite{goa09} (or see Fact 1 and Claim 1 in \cite{jon09} for a self-contained shorter proof) it suffices to establish Theorem \ref{th:main0} for  the scenario $|\mathcal{U}|\leq
|\mathcal{X}|, |\mathcal{V}|\leq |\mathcal{X}|$ and $X=f(U,V)$, a deterministic function of $(U,V)$.
\end{remark}

The outline of the proof is as follows:
\begin{enumerate}
\item We first prove the inequality for some special settings, or ``trivial" cases. (Section~\ref{sec:specset})
\item We show that it suffices to prove  for the nontrivial cases $X=U\land V$ and $X=U\oplus V$.
(Section~\ref{sec:two-cases})
\item For $X=U\land V$, we show that the nontrivial maximum of $LHS$ can only be
achieved when at least two of $\{p_{00},p_{01},p_{10}\}$ equal zero. This reduces the setting to one of the trivial cases. (Section~\ref{sec:u-and-v})
\item For $X=U\oplus V$, we show that the nontrivial maximum of $LHS$ can only
be achieved when at least one of $p_{uv}$ equals zero, which is reduced to the
case $X=U\land V$. (Section~\ref{sec:u-xor-v})
\end{enumerate}  

For a binary-input channel $X\to Y$, let $\{a_i,\hat{a}_i \}$ denote the transition
probabilities, where 
\begin{align*}
\p(Y=i|X=0) = a_i,\quad \p(Y=i|X=1) = \hat{a}_i ,\quad i=1,\dots,N.
\end{align*}
Similarly let
\begin{align*}
\p(Z=i|X=0) = b_i,\quad \p(Z=i|X=1) = \hat{b}_i ,\quad i=1,\dots,N.
\end{align*}

\begin{remark}
W.l.o.g. we can assume that all the terms, $\{a_i,\hat{a}_i, b_i, \hat{b}_i \}$
are non-zero (or in general positive). The validity of the inequality at boundary points, i.e. some of
$\{a_i,\hat{a}_i, b_i, \hat{b}_i \}$ are zero follows from the continuity of  mutual information. 
\label{re:non-zero}
\end{remark}

\smallskip

{\em Notation.} $X\perp Y$: $X$ and $Y$ are independent.

\smallskip

Since $U\to X\to Y$ and $V\to X\to Z$ are Markov chains, from data
processing inequality, we know 
\begin{align}
I(U;Y)\leq I(X;Y),\quad I(U;Y)\leq I(U;X), \nonumber \\
I(V;Z)\leq I(X;Z),\quad I(V;Z)\leq I(V;X). \label{eqn:data-proc}
\end{align}
With these inequalities, we first prove Theorem~\ref{th:main0} for some special settings.

\subsection{Proof for Special Settings}
\label{sec:specset}

\begin{enumerate}
\item[SS1.] $a_i\equiv\hat{a}_i$. Then $X\perp Y$,
and thus $I(U;Y) = I(X;Y)=0$. Thus from \eqref{eqn:data-proc} and the
non-negativity of $I(U;V)$ we have $I(V;Z) - I(V;U) \leq I(X;Z)$, i.e.  Theorem~\ref{th:main0} holds. Similarly Theorem~\ref{th:main0} holds when $b_i\equiv\hat{b}_i$.

\item[SS2.] $U\perp X$. Then $I(U;Y) = I(U;X)=0$. Again from \eqref{eqn:data-proc} and the
non-negativity of $I(U;V)$  Theorem~\ref{th:main0} holds.
Similarly when $V\perp X$, Theorem~\ref{th:main0} also holds.


\end{enumerate}

\subsection{Two Nontrivial Cases} \label{sec:two-cases}
According to Remark \ref{re:suff}, to prove the
inequality~(\ref{eq:main}), it suffices to consider $X=f(U,V)$ with binary
$U$ and $V$. Notice there are 16 possible functions $f$, and they can be
classified into the following equivalent (equivalence is due to relabeling) groups
\begin{enumerate}
 \item[$G_1$:] $X=\{0\}, X=\{1\}$
 \item[$G_2$:] $X=U, X=\bar{U}, X=V, X=\bar{V}$
 \item[$G_3$:] $X=U\land V, X=\bar{U}\land V, X=U\land\bar{V},
X=\bar{U}\land\bar{V}$
 \item[$G_4$:] $X=U\lor V, X=\bar{U}\lor V, X=U\lor\bar{V},
X=\bar{U}\lor\bar{V}$
 \item[$G_5$:] $X=U\oplus V,X=\bar{U}\oplus V$
\end{enumerate}
The reason that these are equivalent groups is that, in each group, all the
cases can be reduced to the first case by using some bijections. For
example, in $G_3$, let the distributions of $(U,V)$ be $p(u,v)$ and $r(u,v)$
for $X=U\land V$ and $X=\bar{U}\land V$, respectively. The bijection is
$p_{00}\leftrightarrow r_{10}$, $p_{01}\leftrightarrow r_{11}$,
$p_{10}\leftrightarrow r_{00}$, $p_{11}\leftrightarrow r_{01}$. Thus, we just
need to prove Theorem~\ref{th:main0} for the first function in each group.

Further, notice for the case $X=U\lor V$ with $q(u,v)$, by bijection
$p_{00}\leftrightarrow q_{11}$, $p_{01}\leftrightarrow q_{01}$,
$p_{10}\leftrightarrow q_{10}$, $p_{11}\leftrightarrow q_{00}$, we can also use
the same proof as for the case $X=U\land V$. That is, we use the fact that $X=U\lor V \Leftrightarrow \bar{X} = \bar{U} \land \bar{V}$ to reduce the proof of the ``or" case of one channel to the ``and" case of another broadcast channel obtained by flipping $U,V$, and $X$.

So it remains to consider the
first cases of the groups except $G_4$.

The first two cases are trivial. For $X=\{0\}$, the theorem is reduced to
$-I(U;V)\leq 0$. For $X=U$, i.e. $I(U;Y) = I(X;Y)$, the theorem follows from the data processing
inequality, $I(V;Z)\leq I(V;U) = I(V;X)$ (see Eqn.(\ref{eqn:data-proc})). So finally we
just need to consider the following two nontrivial cases:
\begin{enumerate}
\item[$C_3$:] $X=U \land V$
\item[$C_5$:] $X=U\oplus V$
\end{enumerate}

\subsection{Proof for Case $X=U \land V$} \label{sec:u-and-v}
In this case, $\p(X=0)=p_{11}$. Now we fix $p_{11}$, the $RHS$ keeps unchanged
with given $Y$ and $Z$. If $p_{11}$ equals to 0 (or 1), then $X=\{0\}$ (or
$X=\{1\}$), and it reduces to the group $G_1$. So we just need to consider
$p_{11}\in(0,1)$. Take $(p_{10},p_{01})$ as the free variables, with
$p_{00}=1-p_{11}-p_{01}-p_{10}$. Thus the region of possible $(p_{10},p_{01})$
is a right triangle containing the interior. The basic idea of the proof is
that:
\begin{enumerate}
\item We first prove Theorem~\ref{th:main0} at the vertices of the region of
$(p_{10},p_{01})$. (Section~\ref{sec:case-and-vertex})
\item Then we show that any nontrivial local maxima of $LHS$ can only be one
vertex of $(p_{10},p_{01})$. (Section~\ref{sec:case-and-edge} and
\ref{sec:case-and-interior})
\end{enumerate}
\subsubsection{Case C3-1: at least two of $p_{00},p_{01},p_{10}=0$}
\label{sec:case-and-vertex}
When this happens, the condition reduces to \{$X=U,V=1$\} or \{$X=V,U=1$\} or
\{$U=V=X$\}; which belong to group $G_2$, where Theorem~\ref{th:main0} holds. Here we
mention that with $p_{11}<1$, these three probabilities cannot be zero
simultaneously. However, for clarity, we still use ``at least two" instead of ``exactly two".

\subsubsection{Case C3-2: exactly one of $p_{00},p_{01},p_{10}=0$}
\label{sec:case-and-edge}
For these cases, we show that nontrivial local maxima does not exist. Consider a
Lyapunov perturbation $q(u,v,x)=p(u,v,x)[1+\varepsilon L(u,v)], \varepsilon\in
\mathcal{R}$ that maintains $\p(X=0)$. This implies that the perturbation
satisfies
\begin{align}
L_{11}=0,\quad p_{00}L_{00}+p_{01}L_{01}+p_{10}L_{10}=0
\label{eq:lpcon1}
\end{align}
For any valid perturbation, at any local maxima of $LHS$, the first and second
derivatives w.r.t. $\varepsilon$ must be $=0$ and $\leq 0$, respectively. Thus
\begin{align} 
H_L(U,V) &=H_{E[L|U,Y]}(U,Y) +H_{E[L|V,Z]}(V,Z)\label{eqn:1st-deri-L}\\
E[E[L|U,V]^2] &\geq E[E[L|U,Y]^2] +E[E[L|V,Z]^2]\label{eqn:2nd-deri-L}
\end{align}
where 
\begin{align*}
H_L(U,V) = &-p_{00}L_{00}\log p_{00}-p_{01}L_{01}\log p_{01}-p_{10}L_{10}\log
p_{10} \\
H_{E[L|U,Y]}(U,Y) = &-\sum a_i(p_{00}L_{00}+p_{01}L_{01})
\log[a_i(p_{00}+p_{01})] \nonumber\\
&-\sum a_ip_{10}L_{10} \log[a_ip_{10}+\hat{a}_i p_{11}] \\
H_{E[L|V,Z]}(V,Z) = &-\sum b_i(p_{00}L_{00}+p_{10}L_{10})
\log[b_i(p_{00}+p_{10})] \nonumber\\
&-\sum b_ip_{01}L_{01} \log[b_ip_{01}+\hat{b}_i p_{11}],
\end{align*}
and
\begin{align*}
E[E[L|U,V]^2] &=p_{00}L_{00}^2 +p_{01}L_{01}^2 +p_{10}L_{10}^2\\
E[E[L|U,Y]^2] &=\frac{(p_{00}L_{00}+p_{01}L_{01})^2} {p_{00}+p_{01}}
+\sum\frac{a_i^2p_{10}^2L_{10}^2}{a_ip_{10}+\hat{a}_i p_{11}}\\
E[E[L|V,Z]^2] &=\frac{(p_{00}L_{00}+p_{10}L_{10})^2} {p_{00}+p_{10}}
+\sum\frac{b_i^2p_{01}^2L_{01}^2}{b_ip_{01}+\hat{b}_i p_{11}}.
\end{align*}

\medskip

{\em Case 1: $p_{00}=0, p_{01},p_{10},p_{11}>0$}

In this case, condition~(\ref{eqn:2nd-deri-L}) implies that the following
inequality holds for all valid perturbations satisfying \eqref{eq:lpcon1}:
\begin{align*}
p_{01}L_{01}^2 +p_{10}L_{10}^2 \geq &+p_{01}L_{01}^2
+\sum\frac{a_i^2p_{10}^2L_{10}^2}{a_ip_{10}+\hat{a}_i p_{11}}\nonumber\\ 
&+p_{10}L_{10}^2 +\sum\frac{b_i^2p_{01}^2L_{01}^2}{b_ip_{01}+\hat{b}_i p_{11}}\\
\Longrightarrow ~0\geq
&\sum\frac{a_i^2p_{10}^2L_{10}^2}{a_ip_{10}+\hat{a}_i p_{11}}
+\sum\frac{b_i^2p_{01}^2L_{01}^2}{b_ip_{01}+\hat{b}_i p_{11}}.
\end{align*}
However, when $p_{01},p_{10},p_{11}>0$,  this cannot hold for all valid
perturbations.


\medskip

{\em Case 2: $p_{01}=0, p_{00},p_{10},p_{11}>0$}

In this case, condition~(\ref{eqn:1st-deri-L}) implies that
\begin{align*}
&\log p_{10}-\log p_{00}=\sum a_i\log[a_ip_{10}+\hat{a}_i p_{11}]-\sum
a_i\log[a_ip_{00}]\\
\Longrightarrow ~&\sum a_i\log[a_ip_{10}] = \sum
a_i\log[a_ip_{10}+\hat{a}_i p_{11}]
\end{align*}
This equality cannot hold since $a_i, \hat{a}_i, p_{10}, p_{11}>0$ (see Remark \ref{re:non-zero}).


\medskip

{\em Case 3: $p_{10}=0, p_{00},p_{01},p_{11}>0$}

Just as in Case 2, condition~(\ref{eqn:1st-deri-L}) implies that 
\begin{align*}
\sum b_i\log[b_ip_{01}] = \sum b_i\log[b_ip_{01}+\hat{b}_i p_{11}]
\end{align*}
As before, this equality cannot hold since $b_i, \hat{b}_i, p_{01}, p_{11}>0$.

\subsubsection{Case C3-3: all $p_{uv}>0$} \label{sec:case-and-interior}
As $\p(X=1)= p_{11}$ (equivalently $H(Y), H(Z)$ via the Markov chain $(U,V) \to X \to (Y,Z)$) is kept fixed, the local maxima of $LHS$ is the same as that of
\begin{align*}
f(p_{10},p_{01}) & = H(U,V)-H(U,Y)-H(V,Z) \\
& = -p_{00}\log p_{00}-p_{01}\log p_{01}-p_{10}\log p_{10}-p_{11}\log
p_{11} \\
& \quad + \sum a_i(p_{00}+p_{01}) \log[a_i(p_{00}+p_{01})] 
 + \sum (a_ip_{10}+\hat{a}_i p_{11}) \log[a_ip_{10}+\hat{a}_i p_{11}] \\
 & \quad + \sum b_i(p_{00}+p_{10}) \log[b_i(p_{00}+p_{10})] 
+\sum (b_ip_{01}+\hat{b}_i p_{11}) \log[b_ip_{01}+\hat{b}_i p_{11}].
\end{align*}

At any local maxima, the gradient $\bold\nabla f$
and Hessian matrix $\bold\nabla^2 f$ must satisfy
\begin{equation}
\bold\nabla f = \vec{0}, ~ \bold\nabla^2 f \preceq \mathbf{0},\
\label{eq:lmax}
\end{equation}
where $\bold\nabla^2 f \preceq \mathbf{0}$ denotes that $\bold\nabla^2 f$ is negative semi-definite. We now compute the gradient and the Hessian to investigate locations of the local maxima.


\medskip

{\em 1. First Derivative}:

 Differentiating w.r.t. the free variables we obtain:
\begin{align*}
\frac{\partial f}{\partial p_{10}} &= \log \frac{p_{00}}{p_{10}} - \sum
a_i\log\frac{a_i(p_{00}+p_{01})}{a_ip_{10}+\hat{a}_i p_{11}}\\
\frac{\partial f}{\partial p_{01}} &= \log \frac{p_{00}}{p_{01}} - \sum
b_i\log\frac{b_i(p_{00}+p_{10})}{b_ip_{01}+\hat{b}_i p_{11}}.
\end{align*}
The condition $\bold\nabla f = \vec{0}$ implies that
\begin{align}
\log\frac{p_{00}}{p_{10}} &= \sum a_i\log \frac{a_i(p_{00}+p_{01})}
{a_ip_{10}+\hat{a}_i p_{11}} \label{eq:dif1}\\
\log\frac{p_{00}}{p_{01}} &= \sum b_i\log \frac{b_i(p_{00}+p_{10})}
{b_ip_{01}+\hat{b}_i p_{11}} \label{eq:dif2}.
\end{align}
Using the concavity of logarithm, we have 
\begin{align}
\frac{p_{00}}{p_{10}} & \leq\sum\frac{a_i^2(p_{00}+p_{01})}
{a_ip_{10}+\hat{a}_i p_{11}} \nonumber\\
\frac{p_{00}}{p_{01}} & \leq\sum\frac{b_i^2(p_{00}+p_{10})}
{b_ip_{01}+\hat{b}_i p_{11}}, \label{eqn:1st-deri-p}
\end{align}
where the equalities hold iff. (using Remark \ref{re:non-zero})
\begin{align*}
a_i\equiv c_a \hat{a}_i, ~ b_i\equiv c_b \hat{b}_i,
\end{align*}
for some constants $c_a, c_b$ respectively.

However since $\sum_i a_i = \sum_i \hat{a}_i = 1$ we obtain that $c_a=1$ (similarly $c_b=1$).
Thus equalities hold iff. 
\begin{equation}
\label{eq:equa}
a_i\equiv \hat{a}_i, ~ b_i\equiv  \hat{b}_i.
\end{equation}

\medskip

\newpage

{\em 2. Second Derivative}:

\smallskip

We now compute the Hessian $G\equiv \bold\nabla^2 f$,
The second derivatives are 
\begin{align*}
G_{11} &= \frac{\partial^2 f}{\partial p_{10}^2} = -\frac{1}{p_{00}}
-\frac{1}{p_{10}} +\frac{1}{p_{00}+p_{01}} +\sum\frac{a_i^2}
{a_ip_{10}+\hat{a}_i p_{11}}\\
G_{12} &= G_{21} = -\frac{1}{p_{00}}\\
G_{22} &= \frac{\partial^2 f}{\partial p_{01}^2} = -\frac{1}{p_{00}}
-\frac{1}{p_{01}} +\frac{1}{p_{00}+p_{10}} +\sum\frac{b_i^2}
{b_ip_{01}+\hat{b}_i p_{11}}.
\end{align*}

As $p_{01}>0$, we have $G_{11}\leq-\frac{1}{p_{00}} -\frac{1}{p_{10}}
+\frac{1}{p_{00}+p_{01}} +\frac{1} {p_{10}}<0$. Similarly we have $G_{22} < 0$. For $G$ with $G_{11}<0$ and $G_{22}<0$ to be negative semi-definite, it is necessary and sufficient that $\mbox{det}(G) \geq 0$.

From \eqref{eq:dif1}  and \eqref{eqn:1st-deri-p} we have 
\begin{align*}
G_{11} & \geq -\frac{1}{p_{00}} -\frac{1}{p_{10}} +\frac{1}{p_{00}+p_{01}} + \frac{p_{00}}{p_{10}(p_{00}+p_{01})} \\
& = - \frac{p_{01}(p_{00}+p_{10})}{p_{00}p_{10}(p_{00}+p_{01})}.
\end{align*}

Similarly from \eqref{eq:dif2} and \eqref{eqn:1st-deri-p} we have
\begin{equation*}
G_{22} \geq - \frac{p_{10}(p_{00}+p_{01})}{p_{00}p_{01}(p_{00}+p_{10})}.
\end{equation*}
It is clear that equalities in the above two inequalities hold iff. \eqref{eq:equa} holds.

Since $G_{11}, G_{22} < 0$ we have 
$$ G_{11} G_{22} \leq  \frac{p_{01}(p_{00}+p_{10})}{p_{00}p_{10}(p_{00}+p_{01})} \cdot \frac{p_{10}(p_{00}+p_{01})}{p_{00}p_{01}(p_{00}+p_{10})} = \frac{1}{p_{00}^2} = G_{12}^2, $$
with equality holding only if \eqref{eq:equa} holds.

Thus $\mbox{det}(G) < 0$ or there is no local minima when all $p_{uv}>0$ unless the channel parameters satisfy \eqref{eq:equa}. However when \eqref{eq:equa} holds, we know that the inequality is true as it corresponds to the special setting SS1. 

This completes the argument that the inequality is indeed true when $X=U \land V$ as we have already shown the validity of the inequality at the vertices of the region defined by $(p_{10},p_{01})$, the possible locations of the local maxima of the $LHS$.

\subsection{Proof for Case $X=U \oplus V$} \label{sec:u-xor-v}

Similar to the ``and" case; we will show that nontrivial local maxima can't be achieved
when all $p_{uv}>0$. And when at least one of $p_{uv}$ equals zero, it reduces
to the case $X=U\land V$.
\subsubsection{Case C5-1: at least one of $p_{uv}=0$}
This case can be reduced to the group $G_3$ or $G_4$, and further reduced to the case
$X=U\land V$. For example, if $p_{01}=0$, $X=U\oplus V$ is a special case of
$X=U \land \bar{V}$. 
\subsubsection{Case C5-2: all $p_{uv}>0$}

Just as in \cite{jon09} we will consider a more general perturbation (see \cite{jon09} for the motivation).

Consider a perturbation $q(u,v,x)=p(u,v,x)+\varepsilon\lambda(u,v,x)$ for some
$\varepsilon>0$. For a valid perturbation, we require that
$\lambda_{001},\lambda_{010},\lambda_{100},\lambda_{111}\geq0$ as the
corresponding $p(u,v,x)$ are zero. Further let us require the perturbation
maintains $\p(X=0)$, that is 
\begin{align}\label{eqn:lambda-maintain-x}
\lambda_{000}+\lambda_{010}+\lambda_{100}+\lambda_{110}&=0\nonumber\\
\lambda_{001}+\lambda_{011}+\lambda_{101}+\lambda_{111}&=0.
\end{align}
For any perturbation that satisfies the above conditions at any local maximum,
it must be true that the first derivative cannot be positive. This implies that 
\begin{align}\label{eqn:1st-deri-lambda-sim}
H_\lambda (U,V) -H_{E[\lambda|U,Y]}(U,Y) -H_{E[\lambda|V,Z]}(V,Z)\leq 0
\end{align}
where 
\begin{align*}
H_\lambda(U,V) = &-(\lambda_{000}+\lambda_{001}) \log p_{00}
-(\lambda_{010}+\lambda_{011})\log p_{01} \nonumber\\
&-(\lambda_{100}+\lambda_{101}) \log p_{10} -(\lambda_{110}+\lambda_{111})\log
p_{11}\\
H_{E[\lambda|U,Y]}(U,Y) = &-\sum [a_i(\lambda_{000}+\lambda_{010})+
\hat{a}_i (\lambda_{001}+\lambda_{011})] \log[a_ip_{00}+\hat{a}_i p_{01}]
\nonumber\\
&-\sum [a_i(\lambda_{100}+\lambda_{110})+
\hat{a}_i (\lambda_{101}+\lambda_{111})] \log[a_ip_{11}+\hat{a}_i p_{10}]\\
H_{E[\lambda|V,Z]}(V,Z) = &-\sum [b_i(\lambda_{000}+\lambda_{100})+
\hat{b}_i (\lambda_{001}+\lambda_{101})] \log[b_ip_{00}+\hat{b}_i p_{10}]
\nonumber\\
&-\sum [b_i(\lambda_{010}+\lambda_{110})+
\hat{b}_i (\lambda_{011}+\lambda_{111})] \log[b_ip_{11}+\hat{b}_i p_{01}].
\end{align*}

From Eqn.(\ref{eqn:lambda-maintain-x}), we express $\lambda_{000}$ and
$\lambda_{011}$ in the term of other $\lambda(u,v,x)$ variables, that is 
\begin{align*}
\lambda_{000} = -\lambda_{010}-\lambda_{100}-\lambda_{110},\\
\lambda_{011} = -\lambda_{001}-\lambda_{101}-\lambda_{111}.
\end{align*}
Substituting the above equations into Eqn.(\ref{eqn:1st-deri-lambda-sim}), we have
\begin{align}\label{eqn:1st-deri-lambda}
&+(\lambda_{010}+\lambda_{100}+\lambda_{110}-\lambda_{001})\log p_{00} 
-(\lambda_{100}+\lambda_{101})\log p_{10} \nonumber\\
&+(\lambda_{001}+\lambda_{101}+\lambda_{111}-\lambda_{010})\log p_{01}
-(\lambda_{110}+\lambda_{111})\log p_{11} \nonumber\\
\leq
&+\sum [a_i(\lambda_{100}+\lambda_{110})+
\hat{a}_i (\lambda_{101}+\lambda_{111})]
\log\frac{a_ip_{00}+\hat{a}_i p_{01}}{a_ip_{11}+ \hat{a}_i p_{10}} \nonumber\\
&+\sum [b_i(\lambda_{010}+\lambda_{110})-
\hat{b}_i (\lambda_{001}+\lambda_{101})]
\log\frac{b_ip_{00}+\hat{b}_i p_{10}}{b_ip_{11}+ \hat{b}_i p_{01}}
\end{align}
Since (\ref{eqn:1st-deri-lambda}) holds for any $\lambda_{110}$
and any nonnegative $\lambda_{010},\lambda_{100}$, it implies  that
\begin{align}
\log\frac{p_{00}}{p_{11}} = &\sum
a_i\log\frac{a_ip_{00}+\hat{a}_i p_{01}}{a_ip_{11}+\hat{a}_i p_{10}} +\sum
b_i\log\frac{b_ip_{00}+\hat{b}_i p_{10}}{b_ip_{11}+\hat{b}_i p_{01}}
\nonumber\\
\log\frac{p_{00}}{p_{01}} \leq &\sum
b_i\log\frac{b_ip_{00}+\hat{b}_i p_{10}}{b_ip_{11}+\hat{b}_i p_{01}}
\label{eq:free1}\\
\log\frac{p_{00}}{p_{10}}\leq& \sum
a_i\log\frac{a_ip_{00}+\hat{a}_i p_{01}}{a_ip_{11}+\hat{a}_i p_{10}}. \label{eq:free2}
\end{align}
These implications come from computing the coefficients of $\lambda_{110}$, $\lambda_{010}$, and $\lambda_{100}$.
The above three equations lead to 
\begin{align}
& \log\frac{p_{00}^2}{p_{01}p_{10}}\leq \log\frac{p_{00}}{p_{11}} \nonumber\\
\Longrightarrow ~& p_{00}p_{11}\leq p_{01}p_{10}. \label{eq:xor1}
\end{align}
Similarly, since the inequality (\ref{eqn:1st-deri-lambda}) also holds for any
$\lambda_{101}$ and any nonnegative $\lambda_{001},\lambda_{111}$, we obtain that
\begin{align}
\log\frac{p_{01}}{p_{10}} =&+\sum
\hat{a}_i \log\frac{a_ip_{00}+\hat{a}_i p_{01}}{a_ip_{11}+\hat{a}_i p_{10}}
-\sum \hat{b}_i \log\frac{b_ip_{00}+\hat{b}_i p_{10}}{b_ip_{11}+\hat{b}_i
p_{01}}\nonumber \\
\log\frac{p_{01}}{p_{00}}\leq& -\sum
\hat{b}_i \log\frac{b_ip_{00}+\hat{b}_i p_{10}}{b_ip_{11}+\hat{b}_i
p_{01}}\label{eq:free3}\\
\log\frac{p_{01}}{p_{11}}\leq& +\sum
\hat{a}_i \log\frac{a_ip_{00}+\hat{a}_i p_{01}}{a_ip_{11}+\hat{a}_i
p_{10}}. \label{eq:free4}
\end{align}
The above three equations lead to 
\begin{align}
& \log\frac{p_{01}^2}{p_{00}p_{11}}\leq \log\frac{p_{01}}{p_{10}} \nonumber \\
\Longrightarrow ~& p_{00}p_{11}\geq p_{01}p_{10}. \label{eq:xor2}
\end{align}

Combining \eqref{eq:xor1} and \eqref{eq:xor2} we obtain that
\begin{align}\label{eqn:p-det-zero}
p_{00}p_{11}=p_{01}p_{10}.
\end{align}
This equality means that the equality holds in \eqref{eq:free1}, \eqref{eq:free2}, \eqref{eq:free3}, and \eqref{eq:free4}. 

In particular, the equalities in \eqref{eq:free1} and \eqref{eq:free3} implies that
\begin{align*}
\log\frac{p_{00}}{p_{01}} =& \sum
b_i\log\frac{b_ip_{00}+\hat{b}_i p_{10}}{b_ip_{11}
+\hat{b}_i p_{01}} = \sum
\hat{b}_i \log\frac{b_ip_{00}+\hat{b}_i p_{10}}{b_ip_{11}+\hat{b}_i p_{01}}.
\end{align*}
Taking a weighted sum, we get
\begin{align}
(p_{00}+p_{10})\log\frac{p_{00}}{p_{01}} =& \sum (b_ip_{00}+\hat{b}_i p_{10})
\log\frac{b_ip_{00}+\hat{b}_i p_{10}}{b_ip_{11}+\hat{b}_i p_{01}}
\label{eqn:kld-equal-b}
\end{align}
From above and using K-L divergence, we have 
\begin{align*}
\log\frac{p_{00}}{p_{01}} =& \sum
\frac{b_ip_{00}+\hat{b}_i p_{10}}{p_{00}+p_{10}}
\log\frac{b_ip_{00}+\hat{b}_i p_{10}}{b_ip_{11}+\hat{b}_i p_{01}} \\
\geq &\log\frac{p_{00}+p_{10}}{p_{11}+p_{01}} =\log\frac{p_{00}}{p_{01}}
\end{align*}
Notice the last equality  holds since $p_{00}p_{11}=p_{01}p_{10}$. Since the K-L divergence inequality is indeed an equality, we require that
\begin{align*}
\frac{b_ip_{00}+\hat{b}_i p_{10}}{b_ip_{11}+\hat{b}_i p_{01}}
\equiv \frac{p_{00}}{p_{01}}.
\end{align*}
From the above we obtain
\begin{align}
(p_{01}-p_{11})(b_{i}-\hat{b}_i )\equiv 0.
\end{align}
Similarly using the fact that we have equalities in \eqref{eq:free2} and \eqref{eq:free4}, we can obtain
\begin{align} (p_{10}-p_{11})(a_{i}-\hat{a}_i )\equiv 0.
\end{align}
Now we have two cases
\begin{enumerate}
\item $b_i\equiv\hat{b}_i $, or $a_i\equiv\hat{a}_i$. In this case the Theorem holds (special setting SS1).
\item $p_{01}=p_{11}, ~p_{10}=p_{11}$. Combining this with $p_{00}p_{11} = p_{01}p_{10}$ (Eqn.\eqref{eqn:p-det-zero}) one obtains that $p_{uv}=1/4$, and as a result $U,V$ and $X$ are mutually independent. The Theorem holds (special setting SS2).
\end{enumerate}
If neither of these two cases is satisfied, there would be no local maxima for $p_{uv}>0$. This shows that the inequality indeed holds when $X=U \oplus V$. This completes the proof of Theorem \ref{th:main0}.

\section{Conclusion}
An information theoretic inequality is established for binary input broadcast channels. This can be used to show that the sum-rate given by Marton's inner bound is indeed equivalent to that given by randomized time-division strategy. 

The proof technique is directly motivated from \cite{jon09} and generalizes the result there.
Clearly the inequality fails when $|X| \geq 3$ (for instance, the Blackwell
channel), so a natural question is whether there is a correct generalization for
higher cardinality input-alphabets. 

It would also be useful to find a more intuitive (geometric) argument to shed more light into the actual counting of the sizes of typical sets. Here is an equivalent formulation which is related to the sizes of certain typical sets.
It can be shown that the information inequality is equivalent to showing that
$$ H(U|Y) + H(V|Z) \geq \min\{H(UV|Y), H(UV|Z)\} $$
whenever $(U,V) \to X \to (Y,Z)$ forms a Markov chain, $X=f(U,V)$ and $|X|=2$.

\section*{Acknowedgements}
The guess that the inequality (Theorem \eqref{th:main0}) may hold in this generality was primarily motivated from another problem that the authors were working with Shlomo Shamai. Indeed the original guess of the authors were that this inequality may hold for binary-input output-symmetric broadcast channels. When the proof of this materialized, the authors realized that they had not used the fact that the outputs needed to be symmetric. Therefore the authors would like to express their thanks to Shlomo Shamai for his part in their work on binary-input output-symmetric broadcast channels.

The authors are also grateful to Raymond Yeung for his insightful comments about the relationship of this inequality to $\Gamma_N^*$.


\bibliographystyle{IEEEtran} \bibliography{mybiblio}

\begin{thebibliography}{1}
\providecommand{\url}[1]{#1}
\csname url@samestyle\endcsname
\providecommand{\newblock}{\relax}
\providecommand{\bibinfo}[2]{#2}
\providecommand{\BIBentrySTDinterwordspacing}{\spaceskip=0pt\relax}
\providecommand{\BIBentryALTinterwordstretchfactor}{4}
\providecommand{\BIBentryALTinterwordspacing}{\spaceskip=\fontdimen2\font plus
\BIBentryALTinterwordstretchfactor\fontdimen3\font minus
  \fontdimen4\font\relax}
\providecommand{\BIBforeignlanguage}[2]{{%
\expandafter\ifx\csname l@#1\endcsname\relax
\typeout{** WARNING: IEEEtran.bst: No hyphenation pattern has been}%
\typeout{** loaded for the language `#1'. Using the pattern for}%
\typeout{** the default language instead.}%
\else
\language=\csname l@#1\endcsname
\fi
#2}}
\providecommand{\BIBdecl}{\relax}
\BIBdecl

\bibitem{cov72}
T.~Cover, ``Broadcast channels,'' \emph{IEEE Trans. Info. Theory}, vol. IT-18,
  pp. 2--14, January, 1972.

\bibitem{cov98}
------, ``Comments on broadcast channels,'' \emph{IEEE Trans. Info. Theory},
  vol. IT-44, pp. 2524--2530, October, 1998.

\bibitem{mar79}
K.~Marton, ``A coding theorem for the discrete memoryless broadcast channel,''
  \emph{IEEE Trans. Info. Theory}, vol. IT-25, pp. 306--311, May, 1979.

\bibitem{goa09}
A.~A. Gohari and V.~Anantharam, ``Evaluation of marton's inner bound for the
  general broadcast channel,'' \emph{CoRR}, vol. abs/0904.4541, 2009.

\bibitem{nae07}
C.~Nair and A.~El~Gamal, ``An outer bound to the capacity region of the
  broadcast channel,'' \emph{IEEE Trans. Info. Theory}, vol. IT-53, pp.
  350--355, January, 2007.

\bibitem{naw08}
C.~Nair and V.~W. Zizhou, ``On the inner and outer bounds for 2-receiver
  discrete memoryless broadcast channels,'' \emph{Proceedings of the ITA
  Workshop}, 2008.

\bibitem{jon09}
\BIBentryALTinterwordspacing
V.~Jog and C.~Nair, ``An information inequality for the bssc channel,'' 2009.
  [Online]. Available:
  \url{http://www.citebase.org/abstract?id=oai:arXiv.org:0901.1492}
\BIBentrySTDinterwordspacing

\end{thebibliography}

\end{document}